\documentstyle[prd,aps,preprint,tighten,epsfig]{revtex}

\begin{document}

\draft

\title{Interference bands in decays of doubly-charged Higgs bosons
to dileptons in the minimal type-II seesaw model at the TeV scale}
\author{{\bf Ping Ren} \thanks{E-mail: renp@ihep.ac.cn}
and {\bf Zhi-zhong Xing}
\thanks{E-mail: xingzz@ihep.ac.cn}}
\address{Institute of High Energy Physics, Chinese Academy of
Sciences, P.O. Box 918, \\
Beijing 100049, China}

\maketitle

\begin{abstract}
The dileptonic decays of doubly-charged Higgs bosons $H^{\pm\pm}$
are investigated in the minimal type-II seesaw model with one Higgs
triplet $\Delta$ and one heavy Majorana neutrino $N^{}_1$ at the TeV
scale. We show that the branching ratios ${\cal B}(H^{\pm\pm}
\rightarrow l^\pm_\alpha l^\pm_\beta)$ depend not only on the mass
and mixing parameters of three light neutrinos $\nu^{}_i$ (for
$i=1,2,3$), but also on those of $N^{}_1$. Assuming the mass of
$N^{}_1$ to lie in the range 200 GeV---1 TeV, we figure out the
generous interference bands for the contributions of $\nu^{}_i$ and
$N^{}_1$ to ${\cal B}(H^{\pm\pm} \rightarrow l^\pm_\alpha
l^\pm_\beta)$: $\sqrt{|\sin\theta^{}_{i4} \sin\theta^{}_{j4}|} \sim
10^{-8}$---$10^{-5}$, where $\theta^{}_{i4}$ and $\theta^{}_{j4}$
measure the strength of charged-current interactions of $N^{}_1$. We
illustrate some salient features of the interference bands by
considering three typical mass patterns of $\nu^{}_i$, and stress
that it is very difficult to distinguish the type-II seesaw model
from the triplet seesaw model in such a parameter region at the
Large Hadron Collider.
\end{abstract}


\section{Introduction}

The effort to build neutrino mass models at the TeV scale has
recently revived \cite{Venice}, simply because this new energy
frontier will soon be explored by the Large Hadron Collider (LHC). A
naive but reasonable argument is that possible new physics, if it
exists at the TeV scale and is responsible for the electroweak
symmetry breaking, might also be responsible for the origin of
neutrino masses. The latter is a kind of new physics which has been
conceivably established by a number of neutrino oscillation
experiments in the past decade \cite{PDG}.

Among many possibilities of generating tiny neutrino masses, a
natural one is to extend the standard model by introducing a few
heavy right-handed Majorana neutrinos \cite{Seesaw1} and (or) one
Higgs triplet \cite{Seesaw2}. The gauge-invariant neutrino mass
terms can then be written as
\begin{eqnarray}
-{\cal L}^{}_{\rm mass} \; = \; \overline{l^{}_{\rm L}} Y^{}_\nu
\tilde{H} N^{}_{\rm R} + \frac{1}{2} \overline{N^{c}_{\rm R}}
M^{}_{\rm R} N^{}_{\rm R} + \frac{1}{2} \overline{l^{}_{\rm L}}
Y^{}_\Delta \Delta i\sigma^{}_2 l^c_{\rm L} + {\rm h.c.} \; ,
%     (1)
\end{eqnarray}
where $M^{}_{\rm R}$ is the mass matrix of right-handed Majorana
neutrinos, and
\begin{equation}
\Delta \; \equiv \; \left(\matrix{H^- & - \sqrt{2} ~ H^0 \cr
\sqrt{2} ~ H^{--} & -H^-}\right)
%     (2)
\end{equation}
denotes the Higgs triplet. After the spontaneous gauge symmetry
breaking, one obtains the neutrino mass matrices $M^{}_{\rm D} =
Y^{}_\nu v/\sqrt{2}$ and $M^{}_{\rm L} = Y^{}_\Delta v^{}_\Delta$,
where $\langle H \rangle \equiv v/\sqrt{2}$ and $\langle \Delta
\rangle \equiv v^{}_\Delta$ correspond to the vacuum expectation
values of the neutral components of $H$ and $\Delta$. To minimize
the degrees of freedom associated with $M^{}_{\rm L}$, $M^{}_{\rm
D}$ and $M^{}_{\rm R}$, we may assume that there is only a single
heavy Majorana neutrino (denoted as $N^{}_1$) in the model. This
assumption implies that $M^{}_{\rm R}$ and $M^{}_{\rm D}$ become
$1\times 1$ and $3\times 1$, respectively, but $M^{}_{\rm L}$
remains to be $3\times 3$. Such a simple seesaw scenario is
phenomenologically viable and can be referred to as the minimal
type-II seesaw model \cite{Gu}. Its simplicity makes it
interesting and instructive to reveal the salient features of the
type-II seesaw mechanism. Therefore, we shall concentrate on this
model in the present paper.

Our purpose is to investigate the dileptonic decays of
doubly-charged Higgs bosons $H^{\pm\pm}$ in the minimal type-II
seesaw model. Such decays can naturally happen because $\Delta$ is
allowed to couple to the standard-model Higgs doublet $H$ and thus
the lepton number is violated by two units \cite{Seesaw2}. If the
mass scale of $\Delta$ is of ${\cal O}(1)$ TeV, then $H^{\pm\pm}$
can be produced at the LHC via the Drell-Yan process $q\bar{q}
\rightarrow \gamma^*, Z^* \rightarrow H^{++} H^{--}$ or through the
charged-current process $q\bar{q}^\prime \rightarrow W^* \rightarrow
H^{\pm\pm}H^{\mp}$. Note that the masses of $H^{\pm\pm}$ and $H^\pm$
are expected to be nearly degenerate in a class of seesaw models
\cite{Seesaw2,Zhou,Triplet}, so only $H^{\pm\pm}\rightarrow
l^{\pm}_\alpha l^{\pm}_\beta$ (for $\alpha, \beta = e, \mu, \tau$)
and $H^{\pm\pm} \rightarrow W^\pm W^\pm$ modes are kinematically
open. Note also that the dileptonic channels $H^{\pm\pm}\rightarrow
l^{\pm}_\alpha l^{\pm}_\beta$ become dominant when $v^{}_\Delta < 1$
MeV is taken \cite{Triplet}. Therefore, we focus our interest on the
same-sign dilepton events of $H^{\pm\pm}$, which signify the lepton
number violation and serve for a clean collider signature of new
physics beyond the standard model \cite{KS}. The rates of
$H^{\pm\pm} \to l^\pm_\alpha l^\pm_\beta$ decays are given by
\begin{equation}
\Gamma(H^{\pm\pm} \to l^\pm_\alpha l^\pm_\beta) \; = \;
\frac{1}{4\pi\left(1+\delta^{}_{\alpha\beta}\right)} ~
|\left(Y^{}_{\Delta}\right)^{}_{\alpha\beta}|^2 M^{}_{H^{\pm\pm}}
\; ,
%     (3)
\end{equation}
from which one obtains the branching ratios \cite{Triplet}
\begin{equation}
{\cal B}(H^{\pm\pm} \to l_\alpha^\pm l_\beta^\pm) \; \equiv \;
\frac{\Gamma(H^{\pm\pm} \to l^\pm_\alpha l^\pm_\beta)}
{\displaystyle\sum^{}_{\rho,\sigma}\Gamma(H^{\pm\pm} \to
l^\pm_\rho l^\pm_\sigma)} =
\frac{2}{\left(1+\delta^{}_{\alpha\beta}\right)} ~
\frac{|\left(M^{}_{\rm L}\right)^{}_{\alpha\beta}|^2}
{\displaystyle\sum^{}_{\rho,\sigma}|\left(M^{}_{\rm
L}\right)^{}_{\rho\sigma}|^2} \; ,
%     (4)
\end{equation}
where the Greek subscripts run over $e$, $\mu$ and $\tau$. It
becomes obvious that the magnitudes of ${\cal B}(H^{\pm\pm} \to
l_\alpha^\pm l_\beta^\pm)$ are only relevant to the matrix elements
of $M^{}_{\rm L}$.

We find that the branching ratios ${\cal B}(H^{\pm\pm}\rightarrow
l^\pm_\alpha l^\pm_\beta)$ depend not only on the masses ($m^{}_1,
m^{}_2, m^{}_3$), flavor mixing angles ($\theta^{}_{12},
\theta^{}_{13}, \theta^{}_{23}$) and CP-violating phases
($\delta^{}_{12}, \delta^{}_{13}, \delta^{}_{23}$) of three light
neutrinos $\nu^{}_1$, $\nu^{}_2$ and $\nu^{}_3$, but also on the
mass ($M^{}_1$) and mixing parameters ($\theta^{}_{14},
\theta^{}_{24}, \theta^{}_{34}$ and $\delta^{}_{14},
\delta^{}_{24}, \delta^{}_{34}$) of the heavy Majorana neutrino
$N^{}_1$. When the former contribution is negligibly small, we can
reproduce the case discussed in Ref. \cite{Zhou}; but when the
contribution of $N^{}_1$ is negligibly small, our results for
${\cal B}(H^{\pm\pm}\rightarrow l^\pm_\alpha l^\pm_\beta)$ can
simply reproduce those obtained in the triplet seesaw model
\cite{LL,Han2}. The new and most interesting case, which has not
been analyzed before, is the competition or interference between
the contributions of light and heavy Majorana neutrinos. Typically
assuming $M^{}_1 \sim 200 ~ {\rm GeV}$---1 TeV and taking three
possible mass patterns of $\nu^{}_i$ as allowed by current
neutrino oscillation data, we figure out the generous interference
bands of $\nu^{}_i$ and $N^{}_1$ contributions to ${\cal
B}(H^{\pm\pm}\rightarrow l^\pm_\alpha l^\pm_\beta)$:
$\sqrt{|\sin\theta^{}_{i4} \sin\theta^{}_{j4}|} \sim
10^{-8}$---$10^{-5}$ (for $i, j =1,2,3$). We stress that both
constructive and destructive interference effects are possible in
this parameter region, in which it is very difficult to
distinguish the type-II seesaw model from the triplet seesaw model
at the LHC. We present some detailed numerical calculations of
${\cal B}(H^{\pm\pm}\rightarrow l^\pm_\alpha l^\pm_\beta)$ in the
interference bands. Although our numerical results are subject to
the minimal type-II seesaw model, they can serve as a good example
to illustrate the interplay between light and heavy Majorana
neutrinos in a generic type-II seesaw scenario.

\section{Interference bands}

After the spontaneous electroweak symmetry breaking, we rewrite
Eq. (1) as
\begin{eqnarray}
-{\cal L}^\prime_{\rm mass} \; = \; \frac{1}{2} ~ \overline{\left(
\nu^{}_{\rm L} ~N^c_{\rm R}\right)} ~ \left( \matrix{ M^{}_{\rm L}
& M^{}_{\rm D} \cr M^T_{\rm D} & M^{}_{\rm R}}\right) \left(
\matrix{ \nu^c_{\rm L} \cr N^{}_{\rm R}}\right) + {\rm h.c.} \; .
%     (5)
\end{eqnarray}
We assume the existence of only a single heavy Majorana neutrino
$N^{}_1$ in the type-II seesaw scenario. The $4\times 4$ neutrino
mass matrix in Eq. (5) is symmetric and can be diagonalized by the
following unitary transformation:
\begin{eqnarray}
\left(\matrix{V & R \cr S & U}\right)^\dagger \left( \matrix{
M^{}_{\rm L} & M^{}_{\rm D} \cr M^T_{\rm D} & M^{}_{\rm R}}\right)
\left(\matrix{V & R \cr S & U}\right)^*  = \left( \matrix{
\widehat{M}^{}_\nu & {\bf 0} \cr {\bf 0} & M^{}_1}\right) \; ,
%     (6)
\end{eqnarray}
where $\widehat{M}^{}_\nu = {\rm Diag}\{m^{}_1, m^{}_2, m^{}_3\}$
with $m^{}_i$ being the masses of three light neutrinos
$\nu^{}_i$, and $M^{}_1$ denotes the mass of $N^{}_1$. After this
diagonalization, the flavor states of three light neutrinos
$\nu^{}_\alpha$ (for $\alpha = e, \mu, \tau$) can be expressed in
terms of the masses states of both three light Majorana neutrinos
$\nu^{}_i$ (for $i=1, 2, 3$) and the heavy Majorana neutrino
$N^{}_1$; namely, $\nu^{}_\alpha = V^{}_{\alpha i} \nu^{}_i +
R^{}_{\alpha 1} N^{}_1$. Then it is straightforward to write out
the standard charged-current interactions between $\nu^{}_\alpha$
and $\alpha$ in the basis of mass states:
\begin{eqnarray}
-{\cal L}^{}_{\rm cc} \; = \; \frac{g}{\sqrt{2}} \left[
\overline{\left(e~~ \mu~~ \tau\right)^{}_{\rm L}} ~V \gamma^\mu
\left( \matrix{\nu^{}_1 \cr \nu^{}_2 \cr \nu^{}_3} \right)^{}_{\rm
L} W^-_{\mu} + \overline{\left(e~~ \mu~~ \tau\right)^{}_{\rm L}}
~R \gamma^\mu {N}^{}_{1\rm L} W^-_\mu \right] + {\rm h.c.} \; .
%     (7)
\end{eqnarray}
We see that $V$ describes the flavor mixing of three light
neutrinos and three charged leptons, while $R$ determines how
strong the heavy Majorana neutrino interacts with three charged
leptons. In other words, $V$ and $R$ are responsible for neutrino
oscillations of $\nu^{}_i$ and collider signatures of $N^{}_1$,
respectively. Note that $V$ itself is not unitary, because $V
V^\dagger + R R^\dagger = {\bf 1}$ holds as a consequence of
unitarity of the $4\times 4$ transformation matrix in Eq. (6). The
correlation between $V$ and $R$ can be parametrized as
\cite{Xing08}
\begin{eqnarray}
V & = & \left( \matrix{ c^{}_{14} & 0 & 0 \cr -\hat{s}^{}_{14}
\hat{s}^*_{24} & c^{}_{24} & 0 \cr - \hat{s}^{}_{14} c^{}_{24}
\hat{s}^*_{34} & - \hat{s}^{}_{24} \hat{s}^*_{34} & c^{}_{34} \cr}
\right) \left( \matrix{ c^{}_{12} c^{}_{13} & \hat{s}^*_{12}
c^{}_{13} & \hat{s}^*_{13} \cr -\hat{s}^{}_{12} c^{}_{23} -
c^{}_{12} \hat{s}^{}_{13} \hat{s}^*_{23} & c^{}_{12} c^{}_{23} -
\hat{s}^*_{12} \hat{s}^{}_{13} \hat{s}^*_{23} & c^{}_{13}
\hat{s}^*_{23} \cr \hat{s}^{}_{12} \hat{s}^{}_{23} - c^{}_{12}
\hat{s}^{}_{13} c^{}_{23} & -c^{}_{12} \hat{s}^{}_{23} -
\hat{s}^*_{12} \hat{s}^{}_{13} c^{}_{23} & c^{}_{13} c^{}_{23}
\cr} \right) \; , \nonumber \\
R & = & \left( \matrix{ \hat{s}^*_{14} \cr c^{}_{14}
\hat{s}^*_{24} \cr c^{}_{14} c^{}_{24} \hat{s}^*_{34} \cr} \right)
\; ,
%     (8)
\end{eqnarray}
where $c^{}_{ij} \equiv \theta^{}_{ij}$, $s^{}_{ij} \equiv
\sin\theta^{}_{ij}$ and $\hat{s}^{}_{ij} \equiv
e^{i\delta^{}_{ij}} s^{}_{ij}$ with $\theta^{}_{ij}$ and
$\delta^{}_{ij}$ (for $1\leq i < j \leq 4$) being the rotation
angles and phase angles, respectively. If the heavy Majorana
neutrino $N^{}_1$ is decoupled (i.e., $\theta^{}_{14} =
\theta^{}_{24} = \theta^{}_{34} = 0$), $V$ will become a unitary
matrix and take the standard form as advocated in Refs.
\cite{PDG,FX01}. Hence non-vanishing $R$ measures the
non-unitarity of $V$.

Now we make use of Eqs. (6) and (8) to reconstruct $M^{}_{\rm L}$,
which determines the branching ratios of $H^{\pm\pm}\rightarrow
l^\pm_\alpha l^\pm_\beta$ decay modes. We obtain
\begin{equation}
M^{}_{\rm L} \; = \; V \widehat{M}^{}_\nu V^T + M^{}_1 R R^T \; .
%     (9)
\end{equation}
Then the explicit expressions of $\left(M^{}_{\rm
L}\right)^{}_{\alpha\beta}$ can be given in terms of the relevant
neutrino masses, mixing angles and CP-violating phases. In view of
current experimental constraints $s^{}_{13} < 0.16$ \cite{Vissani}
and $s^{}_{i4} \lesssim 0.1$ (for $i=1,2,3$) \cite{Antusch}, we
may simplify the exact results of $\left(M^{}_{\rm
L}\right)^{}_{\alpha\beta}$ by taking $c^{}_{13} \approx c^{}_{i4}
\approx 1$. This good approximation allows us to arrive at
\begin{eqnarray}
\left( M^{}_{\rm L} \right)^{}_{ee} & = & m^{}_{1} c^{2}_{12} +
m^{}_{2} \hat{s}^{*2}_{12} + m^{}_{3} \hat{s}^{*2}_{13} + M^{}_1
\hat{s}^{*2}_{14} \; , \nonumber \\
\left( M^{}_{\rm L} \right)^{}_{\mu\mu} & = & m^{}_{1}
\hat{s}^{2}_{12} c^{2}_{23} + m^{}_{2} c^{2}_{12} c^{2}_{23} +
m^{}_{3}
\hat{s}^{*2}_{23} + M^{}_1 \hat{s}^{*2}_{24} \; , \nonumber \\
\left( M^{}_{\rm L} \right)^{}_{\tau\tau} & = & m^{}_{1}
\hat{s}^{2}_{12} \hat{s}^{2}_{23} + m^{}_{2} c^{2}_{12}
\hat{s}^{2}_{23} + m^{}_{3} c^{2}_{23} + M^{}_1 \hat{s}^{*2}_{34}
\; ; \nonumber \\
\left( M^{}_{\rm L} \right)^{}_{e\mu} & = &
-m^{}_{1} c^{}_{12} \hat{s}^{}_{12} c^{}_{23} + m^{}_{2} c^{}_{12}
\hat{s}^{*}_{12} c^{}_{23} + m^{}_{3} \hat{s}^{*}_{13}
\hat{s}^{*}_{23} + M^{}_1
\hat{s}^{*}_{14} \hat{s}^{*}_{24} \; , \nonumber \\
\left( M^{}_{\rm L} \right)^{}_{e\tau} & = & m^{}_{1} c^{}_{12}
\hat{s}^{}_{12} \hat{s}^{}_{23} - m^{}_{2} c^{}_{12}
\hat{s}^{*}_{12} \hat{s}^{}_{23} + m^{}_{3} \hat{s}^{*}_{13}
c^{}_{23} + M^{}_1 \hat{s}^{*}_{14} \hat{s}^{*}_{34} \; , \nonumber \\
\left( M^{}_{\rm L} \right)^{}_{\mu\tau} & = & -m^{}_{1}
\hat{s}^{2}_{12} c^{}_{23} \hat{s}^{}_{23} - m^{}_{2} c^{2}_{12}
c^{}_{23} \hat{s}^{}_{23} + m^{}_{3} c^{}_{23} \hat{s}^{*}_{23} +
M^{}_1 \hat{s}^{*}_{24} \hat{s}^{*}_{34} \; .
%     (10)
\end{eqnarray}
As a consequence,
\begin{eqnarray}
\sum^{}_{\rho,\sigma}|\left(M^{}_{\rm L}\right)^{}_{\rho\sigma}|^2
& = & \left(m^2_1 + m^2_2 + m^2_3\right) + M^2_1 \left( s^2_{14} +
s^2_{24} + s^2_{34} \right)^2 \nonumber \\
&& + 2 m^{}_1 M^{}_1 {\rm Re} \left[ \left( c^{}_{12}
\hat{s}^{}_{14} - \hat{s}^{}_{12} c^{}_{23} \hat{s}^{}_{24} +
\hat{s}^{}_{12} \hat{s}^{}_{23} \hat{s}^{}_{34} \right)^2 \right]
\nonumber \\
&& + 2 m^{}_2 M^{}_1 {\rm Re} \left[ \left( \hat{s}^{*}_{12}
\hat{s}^{}_{14} + c^{}_{12} c^{}_{23} \hat{s}^{}_{24} - c^{}_{12}
\hat{s}^{}_{23} \hat{s}^{}_{34} \right)^2 \right] \nonumber \\
&& + 2 m^{}_3 M^{}_1 {\rm Re} \left[ \left( \hat{s}^{*}_{13}
\hat{s}^{}_{14} + \hat{s}^{*}_{23} \hat{s}^{}_{24} + c^{}_{23}
\hat{s}^{}_{34} \right)^2 \right] \; .
%     (11)
\end{eqnarray}
By combining Eqs. (10) and (11) with Eq. (4), we are then able to
calculate the branching ratios ${\cal B}(H^{\pm\pm} \to
l_\alpha^\pm l_\beta^\pm)$. There are two extreme cases.

(1) If the heavy Majorana neutrino $N^{}_1$ is essentially
decoupled (i.e., $\theta^{}_{i4} \approx 0$ for $i=1,2,3$), the
unitarity of $V$ will be restored. In this case, the results of
${\cal B}(H^{\pm\pm} \to l_\alpha^\pm l_\beta^\pm)$ are the same
as those obtained in the triplet seesaw model \cite{LL,Han2}.

(2) If the contribution of $N^{}_1$ to $(M^{}_{\rm
L})^{}_{\alpha\beta}$ is dominant, one may simplify Eqs. (10) and
(11) by neglecting the terms proportional to $m^{}_i$ (for
$i=1,2,3$). In this case,
\begin{eqnarray}
{\cal B}(H^{\pm\pm} \to e^\pm e^\pm) & \approx &
\frac{s^4_{14}}{\left(s^2_{14} + s^2_{24} + s^2_{34}\right)^2} \;
, \nonumber \\
{\cal B}(H^{\pm\pm} \to \mu^\pm \mu^\pm) & \approx &
\frac{s^4_{24}}{\left(s^2_{14} + s^2_{24} + s^2_{34}\right)^2} \;
, \nonumber \\
{\cal B}(H^{\pm\pm} \to \tau^\pm \tau^\pm) & \approx &
\frac{s^4_{34}}{\left(s^2_{14} + s^2_{24} + s^2_{34}\right)^2} \;
; \nonumber \\
{\cal B}(H^{\pm\pm} \to e^\pm \mu^\pm) & \approx &
\frac{2s^2_{14}s^2_{24}}{\left(s^2_{14} + s^2_{24} +
s^2_{34}\right)^2} \; , \nonumber \\
{\cal B}(H^{\pm\pm} \to e^\pm \tau^\pm) & \approx &
\frac{2s^2_{14}s^2_{34}}{\left(s^2_{14} + s^2_{24} +
s^2_{34}\right)^2} \; , \nonumber \\
{\cal B}(H^{\pm\pm} \to \mu^\pm \tau^\pm) & \approx &
\frac{2s^2_{24}s^2_{34}}{\left(s^2_{14} + s^2_{24} +
s^2_{34}\right)^2} \; ,
%     (12)
\end{eqnarray}
which only rely on the mixing angles $\theta^{}_{i4}$ (for
$i=1,2,3$). Given $s^{}_{14} \approx 0$, possible signatures of
$H^{\pm\pm} \to \mu^\pm \mu^\pm$, $\mu^\pm\tau^\pm$ and
$\tau^\pm\tau^\pm$ modes at the LHC have been analyzed in Ref.
\cite{Zhou}.

Here let us explore the third interesting case, in which the
contributions of $\nu^{}_i$ and $N^{}_1$ to $(M^{}_{\rm
L})^{}_{\alpha\beta}$ are comparable in magnitude and may give rise
to significant interference effects on the branching ratios of
$H^{\pm\pm} \to l_\alpha^\pm l_\beta^\pm$ decays. To be explicit, we
take $\Delta m^2_{21} \sim 8.0 \times 10^{-5} ~ {\rm eV}^2$ and
$|\Delta m^2_{32}| \sim 2.5 \times 10^{-3} ~ {\rm eV}^2$
\cite{Vissani} as the typical inputs and assume $M^{}_1$ to lie in
the range 200 GeV---1 TeV. There are three possible patterns of the
light neutrino mass spectrum: (1) the normal hierarchy: $m^{}_3 \sim
5.1 \times 10^{-2}$ eV, $m^{}_2 \sim 8.9 \times 10^{-3}$ eV, and
$m^{}_1$ is much smaller than $m^{}_2$; (2) the inverted hierarchy:
$m^{}_2 \sim 5.0 \times 10^{-2}$ eV, $m^{}_1 \sim 4.9 \times
10^{-2}$ eV, and $m^{}_3$ is much smaller than $m^{}_1$; (3) the
near degeneracy: $m^{}_1 \sim m^{}_2 \sim m^{}_3 \sim 0.1$ eV to 0.2
eV, which is consistent with the cosmological upper bound $m^{}_1 +
m^{}_2 + m^{}_3 < 0.61$ eV \cite{Vissani}. In each case, the
contributions of  $\nu^{}_i$ and $N^{}_1$ to $(M^{}_{\rm
L})^{}_{\alpha\beta}$ in Eq. (10) will be of the comparable
magnitude if the mixing angles $\theta^{}_{i4}$ satisfy the
following condition
%%%%%%%%%%%%%%%%%%%%%%%
\footnote{Here we have taken account of $\theta^{}_{12} \sim
34^\circ$, $\theta^{}_{13} < 10^\circ$ and $\theta^{}_{23} \sim
45^\circ$ given by a global analysis of current neutrino oscillation
data in the unitary limit of $V$ \cite{Vissani}.}:
%%%%%%%%%%%%%%%%%%%%%%%
\begin{equation}
s^{}_{i4} s^{}_{j4} \; \sim \; \frac{{\rm max}\{m^{}_1, m^{}_2,
m^{}_3\}}{M^{}_1} \; \sim \; 10^{-14} \cdots 10^{-12} \; ,
%     (13)
\end{equation}
where $i,j = 1,2,3$. In view of this rough estimate, which is
essentially compatible with a more careful numerical analysis, we
can generously set $\sqrt{s^{}_{i4} s^{}_{j4}} \sim
10^{-8}$---$10^{-5}$ as the interference bands of ${\cal
B}(H^{\pm\pm} \to l_\alpha^\pm l_\beta^\pm)$ for $M^{}_1 \sim 200$
GeV---1 TeV. Because the CP-violating phases $\delta^{}_{i4}$ are
completely unrestricted, they may cause either constructive or
destructive effects in the interference bands. We shall numerically
calculate ${\cal B}(H^{\pm\pm} \to l_\alpha^\pm l_\beta^\pm)$ in the
subsequent section to illustrate the interference effects for
different patterns of the light neutrino mass hierarchy.

If $M^{}_1 \lesssim {\cal O}(1)$ TeV and the values of $s^{}_{i4}$
lie in the interference bands obtained above, it will be
impossible to produce and observe $N^{}_1$ at the LHC. The reason
is simply that the interaction of $N^{}_1$ with three charged
leptons is too weak to be detected in this parameter space. Given
the integrated luminosity to be $100 ~ {\rm fb}^{-1}$, for
example, the resonant signature of $N^{}_1$ in the channel
$p\bar{p} \to \mu^{\pm} N^{}_1$ with $N^{}_1 \to \mu^{\pm}W^{\mp}$
at the LHC has been analyzed and the sensitivity of the cross
section $\sigma (p\bar{p} \to \mu^{\pm} \mu ^{\pm} W^{\mp})
\approx \sigma(p\bar{p} \to \mu^\pm N^{}_1) {\cal B}(N^{}_1 \to
\mu^{\pm} W^{\mp})$ to the effective mixing parameter
$S^{}_{\mu\mu} \approx s^4_{24}/(s^2_{14} + s^2_{24} + s^2_{34})$
has been examined in Ref. \cite{Han}. It is found that
$S^{}_{\mu\mu} \geq 7.2 \times 10^{-4}$ (or equivalently,
$s^2_{24} \geq 2.1 \times 10^{-3}$ for $s^{}_{14} \sim s^{}_{24}
\sim s^{}_{34}$) is required in order to get a signature at the
$2\sigma$ level for $M^{}_1 \geq 200$ GeV. This result illustrates
that there will be no chance to probe the existence of $N^{}_1$ in
the interference bands at the LHC.

Nevertheless, it is possible to produce $H^{\pm\pm} $ at the LHC
provided $M^{}_{H^{\pm\pm}} \lesssim {\cal O}(1)$ TeV, and it is
also possible to observe the signatures of $H^{\pm\pm} \to
l_\alpha^\pm l_\beta^\pm$ decays \cite{Zhou,Triplet,LL,Han2}. In
this case, however, the measurements of ${\cal B}(H^{\pm\pm} \to
l_\alpha^\pm l_\beta^\pm)$ themselves are very difficult to tell
whether the existence of $H^{\pm\pm}$ is due to a pure triplet
seesaw model or due to a (minimal) type-II seesaw model.

\section{Numerical examples}

For the sake of simplicity, here we take $\theta^{}_{12} = \arctan
(1/\sqrt{2}) \approx 35.3^\circ$, $\theta^{}_{13} = 0^\circ$ and
$\theta^{}_{23} = 45^\circ$, implying that $V$ takes the well-known
tri-bimaximal mixing pattern \cite{TB} in its unitary limit (i.e.,
$\theta^{}_{i4} =0$). In addition, we switch off the CP-violating
phases $\delta^{}_{12}$, $\delta^{}_{13}$ and $\delta^{}_{23}$ so as
to clearly examine the role of new CP-violating phases
$\delta^{}_{i4}$ in ${\cal B}(H^{\pm\pm} \to l_\alpha^\pm
l_\beta^\pm)$. We fix $\Delta m^2_{21} = 8.0 \times 10^{-5} ~ {\rm
eV}^2$, $|\Delta m^2_{32}| = 2.5 \times 10^{-3} ~ {\rm eV}^2$ and
$M^{}_1 = 500$ GeV in our numerical calculations. To further reduce
the number of free parameters, we shall consider two special cases
for the mixing angles $\theta^{}_{i4}$: (a) $\theta^{}_{14} =
\theta^{}_{24} = \theta^{}_{34}$ and (b) $\theta^{}_{14} = 0$ and
$\theta^{}_{24} = \theta^{}_{34}$; and two special cases for the
CP-violating phases $\delta^{}_{i4}$: (a) $\delta^{}_{14} =
\delta^{}_{24} = \delta^{}_{34} =0$ and (b) $\delta^{}_{14} =
\delta^{}_{24} = \delta^{}_{34} =\pi/2$. Our discussions can be
classified into three parts according to three possible patterns of
the light neutrino mass hierarchy.

\subsection{Normal hierarchy}

We simply take $m^{}_1 =0$, such that $m^{}_2 \approx 8.9 \times
10^{-3}$ eV and $m^{}_3 \approx 5.1 \times 10^{-2}$ eV can be
extracted from the given values of $\Delta m^2_{21}$ and $|\Delta
m^2_{32}|$. For chosen values of $\theta^{}_{12}$, $\theta^{}_{13}$,
$\theta^{}_{23}$ and $\delta^{}_{12}$, $\delta^{}_{13}$,
$\delta^{}_{23}$, Eqs. (10) and (11) can now be simplified to
\begin{eqnarray}
\left( M^{}_{\rm L} \right)^{}_{ee} & = &
\frac{1}{3} m^{}_{2} + M^{}_1 \hat{s}^{*2}_{14} \; , \nonumber \\
\left( M^{}_{\rm L} \right)^{}_{\mu\mu} & = & \frac{1}{3} m^{}_{2} +
\frac{1}{2} m^{}_{3} + M^{}_1 \hat{s}^{*2}_{24} \; , \nonumber \\
\left( M^{}_{\rm L} \right)^{}_{\tau\tau} & = & \frac{1}{3} m^{}_{2}
+ \frac{1}{2} m^{}_{3} + M^{}_1 \hat{s}^{*2}_{34} \; ; \nonumber \\
\left( M^{}_{\rm L} \right)^{}_{e\mu} & = & \frac{1}{3} m^{}_{2} +
M^{}_1 \hat{s}^{*}_{14} \hat{s}^{*}_{24} \; , \nonumber \\
\left( M^{}_{\rm L} \right)^{}_{e\tau} & = & - \frac{1}{3} m^{}_{2}
+ M^{}_1 \hat{s}^{*}_{14} \hat{s}^{*}_{34} \; , \nonumber \\
\left( M^{}_{\rm L} \right)^{}_{\mu\tau} & = & - \frac{1}{3}
m^{}_{2} + \frac{1}{2} m^{}_{3} + M^{}_1 \hat{s}^{*}_{24}
\hat{s}^{*}_{34} \; ,
%     (14)
\end{eqnarray}
and
\begin{eqnarray}
\sum^{}_{\rho,\sigma}|\left(M^{}_{\rm L}\right)^{}_{\rho\sigma}|^2 &
= & \left(m^2_2 + m^2_3\right) + M^2_1 \left( s^2_{14} +
s^2_{24} + s^2_{34} \right)^2 \nonumber \\
&& + \frac{2}{3} m^{}_2 M^{}_1 {\rm Re} \left[ \left(
\hat{s}^{}_{14} + \hat{s}^{}_{24} - \hat{s}^{}_{34} \right)^2
\right] + m^{}_3 M^{}_1 {\rm Re} \left[ \left( \hat{s}^{}_{24} +
\hat{s}^{}_{34} \right)^2 \right] \; .
%     (15)
\end{eqnarray}
Our numerical results for the branching ratios ${\cal B}(H^{\pm\pm}
\to l_\alpha^\pm l_\beta^\pm)$ are shown in FIG. 1. Some comments
and discussions are in order.

FIG. 1(a) is obtained by taking $\theta^{}_{14} = \theta^{}_{24} =
\theta^{}_{34} \equiv \theta$ and $\delta^{}_{14} = \delta^{}_{24} =
\delta^{}_{34} =0$. We see that ${\cal B}(H^{\pm\pm} \to e^\pm
\mu^\pm)$ and ${\cal B}(H^{\pm\pm} \to e^\pm \tau^\pm)$ are
approximately equal beyond the interference band ($3\times 10^{-7}
\lesssim \theta \lesssim 2 \times 10^{-6}$), but their near
degeneracy is lifted in the interference band. In contrast, ${\cal
B}(H^{\pm\pm} \to \mu^\pm \mu^\pm) = {\cal B}(H^{\pm\pm} \to
\tau^\pm \tau^\pm)$ holds in the whole parameter space.

FIG. 1(b) is obtained by taking $\theta^{}_{14} = \theta^{}_{24} =
\theta^{}_{34} \equiv \theta$ and $\delta^{}_{14} = \delta^{}_{24} =
\delta^{}_{34} =\pi/2$. One can see more obvious interference
effects for $\theta$ changing from $10^{-7}$ to $10^{-6}$. In
particular, ${\cal B}(H^{\pm\pm} \to e^\pm \tau^\pm)$ is strongly
enhanced, while ${\cal B}(H^{\pm\pm} \to \mu^\pm \mu^\pm)$, ${\cal
B}(H^{\pm\pm} \to \mu^\pm \tau^\pm)$ and ${\cal B}(H^{\pm\pm} \to
\tau^\pm \tau^\pm)$ are strongly suppressed at $\theta \sim 2\times
10^{-7}$.

FIG. 1(c) is obtained by taking $\theta^{}_{14} = 0$,
$\theta^{}_{24} = \theta^{}_{34} \equiv \theta$ and $\delta^{}_{24}
= \delta^{}_{34} =0$. In this case, there is little interference
between the contributions of $\nu^{}_i$ and $N^{}_1$ to  ${\cal
B}(H^{\pm\pm} \to l_\alpha^\pm l_\beta^\pm)$. It is straightforward
to observe that ${\cal B}(H^{\pm\pm} \to e^\pm e^\pm)$, ${\cal
B}(H^{\pm\pm} \to e^\pm \mu^\pm)$ and ${\cal B}(H^{\pm\pm} \to e^\pm
\tau^\pm)$ are considerably suppressed due to the vanishing of
$\theta^{}_{14}$.

FIG. 1(d) is obtained by taking $\theta^{}_{14} = 0$,
$\theta^{}_{24} = \theta^{}_{34} \equiv \theta$ and
$\delta^{}_{24} = \delta^{}_{34} =\pi/2$. In this case, all the
decay modes involve significant interference effects around
$\theta \sim 2 \times 10^{-7}$. Note that ${\cal B}(H^{\pm\pm} \to
\mu^\pm \tau^\pm)$ undergoes both a minimum and a maximum, which
result from the minimums of its numerator and denominator,
respectively. So do ${\cal B}(H^{\pm\pm} \to \mu^\pm \mu^\pm)$ and
${\cal B}(H^{\pm\pm} \to \tau^\pm \tau^\pm)$. In comparison, the
branching ratio of $H^{\pm\pm} \to e^\pm e^\pm$, $e^\pm \mu^\pm$
or $e^\pm \tau^\pm$ only undergoes a maximum, because its
numerator does not have an appreciable minimum in the interference
band.

\subsection{Inverted hierarchy}

We simply take $m^{}_3 =0$, such that $m^{}_1 \approx 4.9 \times
10^{-2}$ eV and $m^{}_2 \approx 5.0 \times 10^{-2}$ eV can be
extracted from the given values of $\Delta m^2_{21}$ and $|\Delta
m^2_{32}|$. For chosen values of $\theta^{}_{12}$, $\theta^{}_{13}$,
$\theta^{}_{23}$ and $\delta^{}_{12}$, $\delta^{}_{13}$,
$\delta^{}_{23}$, Eqs. (10) and (11) can now be simplified to
\begin{eqnarray}
\left( M^{}_{\rm L} \right)^{}_{ee} & = & \frac{2}{3} m^{}_{1} +
\frac{1}{3} m^{}_{2} + M^{}_1 \hat{s}^{*2}_{14} \; , \nonumber \\
\left( M^{}_{\rm L} \right)^{}_{\mu\mu} & = & \frac{1}{6} m^{}_{1} +
\frac{1}{3} m^{}_{2} + M^{}_1 \hat{s}^{*2}_{24} \; , \nonumber \\
\left( M^{}_{\rm L} \right)^{}_{\tau\tau} & = & \frac{1}{6} m^{}_{1}
+ \frac{1}{3} m^{}_{2} + M^{}_1 \hat{s}^{*2}_{34} \; ; \nonumber \\
\left( M^{}_{\rm L} \right)^{}_{e\mu} & = & \frac{1}{3}
\left(m^{}_{2} - m^{}_{1}\right) + M^{}_1 \hat{s}^{*}_{14} \hat{s}^{*}_{24} \; ,
\nonumber \\
\left( M^{}_{\rm L} \right)^{}_{e\tau} & = & \frac{1}{3}
\left(m^{}_{1} - m^{}_{2}\right) + M^{}_1 \hat{s}^{*}_{14} \hat{s}^{*}_{34} \; ,
\nonumber \\
\left( M^{}_{\rm L} \right)^{}_{\mu\tau} & = & -\frac{1}{6} m^{}_{1}
- \frac{1}{3} m^{}_{2} + M^{}_1 \hat{s}^{*}_{24} \hat{s}^{*}_{34} \;
,
%     (16)
\end{eqnarray}
and
\begin{eqnarray}
\sum^{}_{\rho,\sigma}|\left(M^{}_{\rm L}\right)^{}_{\rho\sigma}|^2 &
= & \left(m^2_1 + m^2_2\right) + M^2_1 \left( s^2_{14} +
s^2_{24} + s^2_{34} \right)^2 \nonumber \\
&& + \frac{1}{3} m^{}_1 M^{}_1 {\rm Re} \left[ \left( 2
\hat{s}^{}_{14} - \hat{s}^{}_{24} + \hat{s}^{}_{34} \right)^2\right]
+ \frac{2}{3} m^{}_2 M^{}_1 {\rm Re} \left[ \left( \hat{s}^{}_{14} +
\hat{s}^{}_{24} - \hat{s}^{}_{34} \right)^2 \right] \; .
%     (17)
\end{eqnarray}
As a consequence of $m^{}_1 \approx m^{}_2$, the contributions of
$\nu^{}_1$ and $\nu^{}_2$ are approximately canceled in $(M^{}_{\rm
L})^{}_{e\mu}$ and $(M^{}_{\rm L})^{}_{e\tau}$. Our numerical
results for the branching ratios ${\cal B}(H^{\pm\pm} \to
l_\alpha^\pm l_\beta^\pm)$ are shown in FIG. 2. Some comments and
discussions are in order.

FIG. 2(a) is obtained by taking $\theta^{}_{14} = \theta^{}_{24} =
\theta^{}_{34} \equiv \theta$ and $\delta^{}_{14} = \delta^{}_{24} =
\delta^{}_{34} =0$. We see that ${\cal B}(H^{\pm\pm} \to e^\pm
\mu^\pm)$ and ${\cal B}(H^{\pm\pm} \to e^\pm \tau^\pm)$ are
essentially degenerate in the whole parameter space, so are ${\cal
B}(H^{\pm\pm} \to \mu^\pm \mu^\pm)$ and ${\cal B}(H^{\pm\pm} \to
\tau^\pm \tau^\pm)$. Different from other branching ratios, ${\cal
B}(H^{\pm\pm} \to \mu^\pm \tau^\pm)$ undergoes a minimum just
because of the minimum of $|(M^{}_{\rm L})^{}_{\mu\tau}|$ at $\theta
\sim 2\times 10^{-7}$.

FIG. 2(b) is obtained by taking $\theta^{}_{14} = \theta^{}_{24} =
\theta^{}_{34} \equiv \theta$ and $\delta^{}_{14} = \delta^{}_{24}
= \delta^{}_{34} =\pi/2$. In this case, the contribution of
$N^{}_1$ to ${\cal B}(H^{\pm\pm} \to l_\alpha^\pm l_\beta^\pm)$
flips the sign such that ${\cal B}(H^{\pm\pm} \to \mu^\pm
\tau^\pm)$ undergoes a maximum because of the minimum in its
denominator. Due to the appearance of a minimum in its numerator,
the branching ratio of $H^{\pm\pm} \to e^\pm e^\pm$, $\mu^\pm
\mu^\pm$ or $\tau^\pm \tau^\pm$ undergoes a minimum when $\theta$
varies in the interference band.

FIG. 2(c) is obtained by taking $\theta^{}_{14} = 0$,
$\theta^{}_{24} = \theta^{}_{34} \equiv \theta$ and $\delta^{}_{24}
= \delta^{}_{34} =0$. In this case, the contributions of $N^{}_1$ to
${\cal B}(H^{\pm\pm} \to e^\pm e^\pm)$, ${\cal B}(H^{\pm\pm} \to
e^\pm \mu^\pm)$ and ${\cal B}(H^{\pm\pm} \to e^\pm \tau^\pm)$ are
vanishing as a consequence of $\theta^{}_{14} = 0$. Hence ${\cal
B}(H^{\pm\pm} \to e^\pm \mu^\pm)$ and ${\cal B}(H^{\pm\pm} \to e^\pm
\tau^\pm)$ are strongly suppressed in the whole parameter space, so
is ${\cal B}(H^{\pm\pm} \to e^\pm e^\pm)$ for $\theta > 10^{-6}$.

FIG. 2(d) is obtained by taking $\theta^{}_{14} = 0$,
$\theta^{}_{24} = \theta^{}_{34} \equiv \theta$ and $\delta^{}_{24}
= \delta^{}_{34} =\pi/2$. We see that the results of ${\cal
B}(H^{\pm\pm} \to e^\pm e^\pm)$, ${\cal B}(H^{\pm\pm} \to e^\pm
\mu^\pm)$ and ${\cal B}(H^{\pm\pm} \to e^\pm \tau^\pm)$ in this case
are essentially the same as those in FIG. 2(c). Because the
contribution of $N^{}_1$ flips the sign, now ${\cal B}(H^{\pm\pm}
\to \mu^\pm \mu^\pm) = {\cal B}(H^{\pm\pm} \to \tau^\pm \tau^\pm)$
undergoes a minimum while ${\cal B}(H^{\pm\pm} \to \mu^\pm
\tau^\pm)$ undergoes a maximum in the interference band.

\subsection{Near degeneracy}

We assume $m^{}_{1}\approx m^{}_{2}\approx m^{}_{3}\approx 0.1 ~
{\rm eV}$. Then $m^{}_2 - m^{}_1 \approx 4.0 \times 10^{-4}$ eV
and $m^{}_3 - m^{}_2 \approx \pm 1.25 \times 10^{-2}$ eV can be
extracted from given values of $\Delta m^2_{21}$ and $|\Delta
m^2_{32}|$, respectively. For chosen values of $\theta^{}_{12}$,
$\theta^{}_{13}$, $\theta^{}_{23}$ and $\delta^{}_{12}$,
$\delta^{}_{13}$, $\delta^{}_{23}$, Eqs. (10) and (11) can now be
simplified to
\begin{eqnarray}
\left( M^{}_{\rm L} \right)^{}_{ee} & \approx & m^{}_1 +
M^{}_1 \hat{s}^{*2}_{14} \; , \nonumber \\
\left( M^{}_{\rm L} \right)^{}_{\mu\mu} & \approx & m^{}_1 +
\frac{1}{2} \left(m^{}_3 - m^{}_2\right) +
M^{}_1 \hat{s}^{*2}_{24} \; , \nonumber \\
\left( M^{}_{\rm L} \right)^{}_{\tau\tau} & \approx & m^{}_1 +
\frac{1}{2} \left(m^{}_3 - m^{}_2\right) +
M^{}_1 \hat{s}^{*2}_{34} \; ; \nonumber \\
\left( M^{}_{\rm L} \right)^{}_{e\mu} & \approx & \frac{1}{3}
(m^{}_{2}-m^{}_{1}) + M^{}_1 \hat{s}^{*}_{14} \hat{s}^{*}_{24} \; , \nonumber \\
\left( M^{}_{\rm L} \right)^{}_{e\tau} & \approx & \frac{1}{3}
(m^{}_{1} -m^{}_{2}) + M^{}_1 \hat{s}^{*}_{14} \hat{s}^{*}_{34} \; , \nonumber \\
\left( M^{}_{\rm L} \right)^{}_{\mu\tau} & \approx & \frac{1}{2} (
m^{}_{3} - m^{}_{2}) + M^{}_1 \hat{s}^{*}_{24} \hat{s}^{*}_{34} \; ,
%     (18)
\end{eqnarray}
where we have neglected the small terms proportional to $m^{}_2 -
m^{}_1$ in $(M^{}_{\rm L})^{}_{ee}$, $(M^{}_{\rm L})^{}_{\mu\mu}$,
$(M^{}_{\rm L})^{}_{\mu\tau}$ and  $(M^{}_{\rm L})^{}_{\tau\tau}$.
In addition,
\begin{eqnarray}
\sum^{}_{\rho,\sigma}|\left(M^{}_{\rm L}\right)^{}_{\rho\sigma}|^2 &
\approx & 3m^2_1 + M^2_1 \left( s^2_{14} +
s^2_{24} + s^2_{34} \right)^2 \nonumber \\
&& + \frac{1}{3} m^{}_1 M^{}_1 {\rm Re} \left[ \left( 2
\hat{s}^{}_{14} - \hat{s}^{}_{24} + \hat{s}^{}_{34} \right)^2 +
2\left( \hat{s}^{}_{14} + \hat{s}^{}_{24} - \hat{s}^{}_{34}
\right)^2 + 3\left( \hat{s}^{}_{24} + \hat{s}^{}_{34} \right)^2
\right] ,
%     (19)
\end{eqnarray}
where we have omitted the small mass differences of $\nu^{}_i$. We
fix $m^{}_3 > m^{}_2$ in our numerical calculations. The results for
the branching ratios ${\cal B}(H^{\pm\pm} \to l_\alpha^\pm
l_\beta^\pm)$ are shown in FIG. 3. Some comments and discussions are
in order.

FIG. 3(a) is obtained by taking $\theta^{}_{14} = \theta^{}_{24} =
\theta^{}_{34} \equiv \theta$ and $\delta^{}_{14} = \delta^{}_{24} =
\delta^{}_{34} =0$. In this case, the near degeneracy of ${\cal
B}(H^{\pm\pm} \to e^\pm \mu^\pm)$, ${\cal B}(H^{\pm\pm} \to e^\pm
\tau^\pm)$ and ${\cal B}(H^{\pm\pm} \to \mu^\pm \tau^\pm)$ is just
because of the smallness of $m^{}_2 - m^{}_1$ and $m^{}_3 - m^{}_2$.
A small discrepancy between ${\cal B}(H^{\pm\pm} \to e^\pm e^\pm)$
and ${\cal B}(H^{\pm\pm} \to \mu^\pm \mu^\pm) = {\cal B}(H^{\pm\pm}
\to \tau^\pm \tau^\pm)$ for $\theta < 7 \times 10^{-7}$ is due to
the small terms proportional to $m^{}_3 - m^{}_2$ in $(M^{}_{\rm
L})^{}_{\mu\mu}$ and $(M^{}_{\rm L})^{}_{\tau\tau}$.

FIG. 3(b) is obtained by taking $\theta^{}_{14} = \theta^{}_{24} =
\theta^{}_{34} \equiv \theta$ and $\delta^{}_{14} = \delta^{}_{24}
= \delta^{}_{34} =\pi/2$. We see some mild interference effects in
all the decay channels. Among them, the branching ratio of
$H^{\pm\pm} \to e^\pm \mu^\pm$, $e^\pm \tau^\pm$ or $\mu^\pm
\tau^\pm$ undergoes a maximum, while the branching ratio of
$H^{\pm\pm} \to e^\pm e^\pm$, $\mu^\pm \mu^\pm$ or $\tau^\pm
\tau^\pm$ undergoes a minimum.

FIG. 3(c) is obtained by taking $\theta^{}_{14} = 0$,
$\theta^{}_{24} = \theta^{}_{34} \equiv \theta$ and
$\delta^{}_{24} = \delta^{}_{34} =0$. In this case, ${\cal
B}(H^{\pm\pm} \to e^\pm \mu^\pm)$ and ${\cal B}(H^{\pm\pm} \to
e^\pm \tau^\pm)$ are strongly suppressed in the whole parameter
space. We see no obvious interference in other decay modes.

FIG. 3(d) is obtained by taking $\theta^{}_{14} = 0$,
$\theta^{}_{24} = \theta^{}_{34} \equiv \theta$ and $\delta^{}_{24}
= \delta^{}_{34} =\pi/2$. One can see that ${\cal B}(H^{\pm\pm} \to
e^\pm e^\pm)$ undergoes a maximum in the interference band, so does
${\cal B}(H^{\pm\pm} \to \mu^\pm \tau^\pm)$. In comparison, ${\cal
B}(H^{\pm\pm} \to \mu^\pm \mu^\pm)={\cal B}(H^{\pm\pm} \to \tau^\pm
\tau^\pm)$ undergoes a minimum. The interference effects in this
case are more significant than those in FIG. 3(b).

\section{Summary}

We have studied the dileptonic decays of doubly-charged Higgs bosons
$H^{\pm\pm}$ in the minimal type-II seesaw model with only one heavy
Majorana neutrino and one Higgs triplet. Their branching ratios
${\cal B}(H^{\pm\pm}\rightarrow l^\pm_\alpha l^\pm_\beta)$ depend
not only on the masses, flavor mixing angles and CP-violating phases
of three light neutrinos $\nu^{}_i$ (for $i=1,2,3$), but also on the
mass ($M^{}_1$) and mixing parameters ($\theta^{}_{i4}$ and
$\delta^{}_{i4}$) of the heavy Majorana neutrino $N^{}_1$. We have
focused our attention on the interference bands of ${\cal
B}(H^{\pm\pm}\rightarrow l^\pm_\alpha l^\pm_\beta)$, in which the
contributions of $\nu^{}_i$ and $N^{}_1$ are comparable in
magnitude. Assuming $M^{}_1 \sim 200 ~ {\rm GeV}$---1 TeV and taking
three possible mass patterns of $\nu^{}_i$ as allowed by current
neutrino oscillation data, we have figured out the generous
interference bands $\sqrt{|\sin\theta^{}_{i4} \sin\theta^{}_{j4}|}
\sim 10^{-8}$---$10^{-5}$ (for $i, j =1,2,3$) and presented a
detailed numerical analysis of ${\cal B}(H^{\pm\pm}\rightarrow
l^\pm_\alpha l^\pm_\beta)$.

We stress that both constructive and destructive interference
effects are possible in the interference bands of ${\cal
B}(H^{\pm\pm}\rightarrow l^\pm_\alpha l^\pm_\beta)$, and thus it is
very difficult to distinguish the (minimal) type-II seesaw model
from the triplet seesaw model in this parameter space. Although our
numerical results are subject to a simplified type-II seesaw
scenario, they can serve as a good example to illustrate the
interplay between light and heavy Majorana neutrinos in a generic
type-II seesaw framework. The latter involves more free parameters,
so the corresponding interference bands of ${\cal
B}(H^{\pm\pm}\rightarrow l^\pm_\alpha l^\pm_\beta)$ will be in a
mess.

It is worth pointing out that the lepton-number-violating decays
of singly-charged Higgs bosons $H^{\pm}$ are also important for
testing the gauge triplet nature of the Higgs field. For example,
the observation of $H^+ \to l^+_\alpha \bar{\nu}^{}_\alpha$ and
$H^- \to l^-_\alpha \nu^{}_\alpha$ (for $\alpha = e, \mu, \tau$)
decays will be particularly useful to determine the mass spectrum
of three light Majorana neutrinos \cite{Han2} because these
processes are independent of the unknown Majorana phases in the
triplet seesaw model. A similar study of the
lepton-number-violating $H^{\pm}$ decays can be done in the
type-II seesaw model, where heavy Majorana neutrinos exist,
although the interference bands of ${\cal B}(H^+ \to l^+_\alpha
\bar{\nu}^{}_\alpha)$ and ${\cal B}(H^- \to l^-_\alpha
\nu^{}_\alpha)$ are expected to be different from those of ${\cal
B}(H^{\pm\pm}\rightarrow l^\pm_\alpha l^\pm_\beta)$. We shall
carry out a systematic analysis of both $H^{\pm\pm}$ decays and
$H^\pm$ decays in the minimal type-II seesaw scenario elsewhere
\cite{Chao}.

It is certainly a big challenge to identify the unique or correct
seesaw mechanism of neutrino mass generation, if such a mechanism
really exists, at the upcoming LHC and the future International
Linear Collider. In particular, the collider signatures of both the
Higgs triplet and heavy Majorana neutrinos will have to be
experimentally established before a claim of having verified the
type-II seesaw mechanism can be made. While the running of the LHC
itself might be very difficult to help us pin down the true flavor
dynamics of leptons and quarks, we hope that it would at least shed
light on what this dynamics looks like at the TeV energy scale.

\vspace{0.5cm}

One of us (Z.Z.X.) is grateful to W. Chao and S. Zhou for helpful
discussions. This work was supported in part by the National
Natural Science Foundation of China.

%%%%%%%%%%%%%%%%%  figure 1      %%%%%%%%%%%%%%%%%%%%%%%%%%%%%%%%%%%%
\begin{figure}
\psfig{file=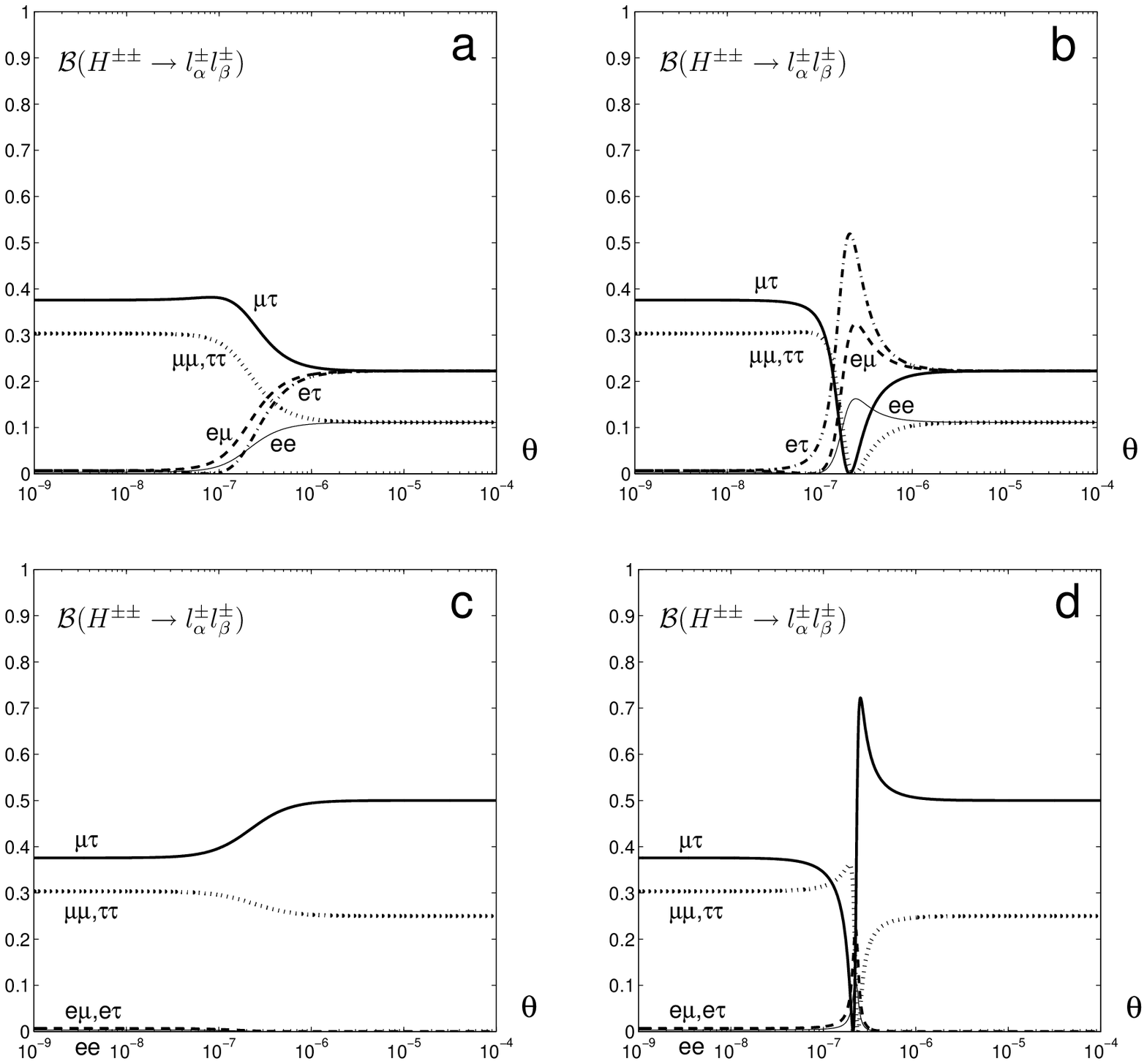, bbllx=2.5cm, bblly=19.6cm, bburx=6.6cm,
bbury=24.0cm, width=3.8cm, height=4cm, angle=0, clip=0}
\vspace{13.5cm} \caption{Branching ratios of $H^{\pm\pm} \to
l^\pm_\alpha l^\pm_\beta$ decays for the normal hierarchy of
$m^{}_i$ with $m^{}_1 =0$: (a) $\theta^{}_{14} = \theta^{}_{24} =
\theta^{}_{34} \equiv \theta$ and $\delta^{}_{14} = \delta^{}_{24} =
\delta^{}_{34} =0$; (b) $\theta^{}_{14} = \theta^{}_{24} =
\theta^{}_{34} \equiv \theta$ and $\delta^{}_{14} = \delta^{}_{24} =
\delta^{}_{34} =\pi/2$;  (c) $\theta^{}_{14} = 0$, $\theta^{}_{24} =
\theta^{}_{34} \equiv \theta$ and $\delta^{}_{24} = \delta^{}_{34}
=0$; (d) $\theta^{}_{14} = 0$, $\theta^{}_{24} = \theta^{}_{34}
\equiv \theta$ and $\delta^{}_{24} = \delta^{}_{34} =\pi/2$. }
\end{figure}

%%%%%%%%%%%%%%%%%  figure 2      %%%%%%%%%%%%%%%%%%%%%%%%%%%%%%%%%%%%
\begin{figure}
\psfig{file=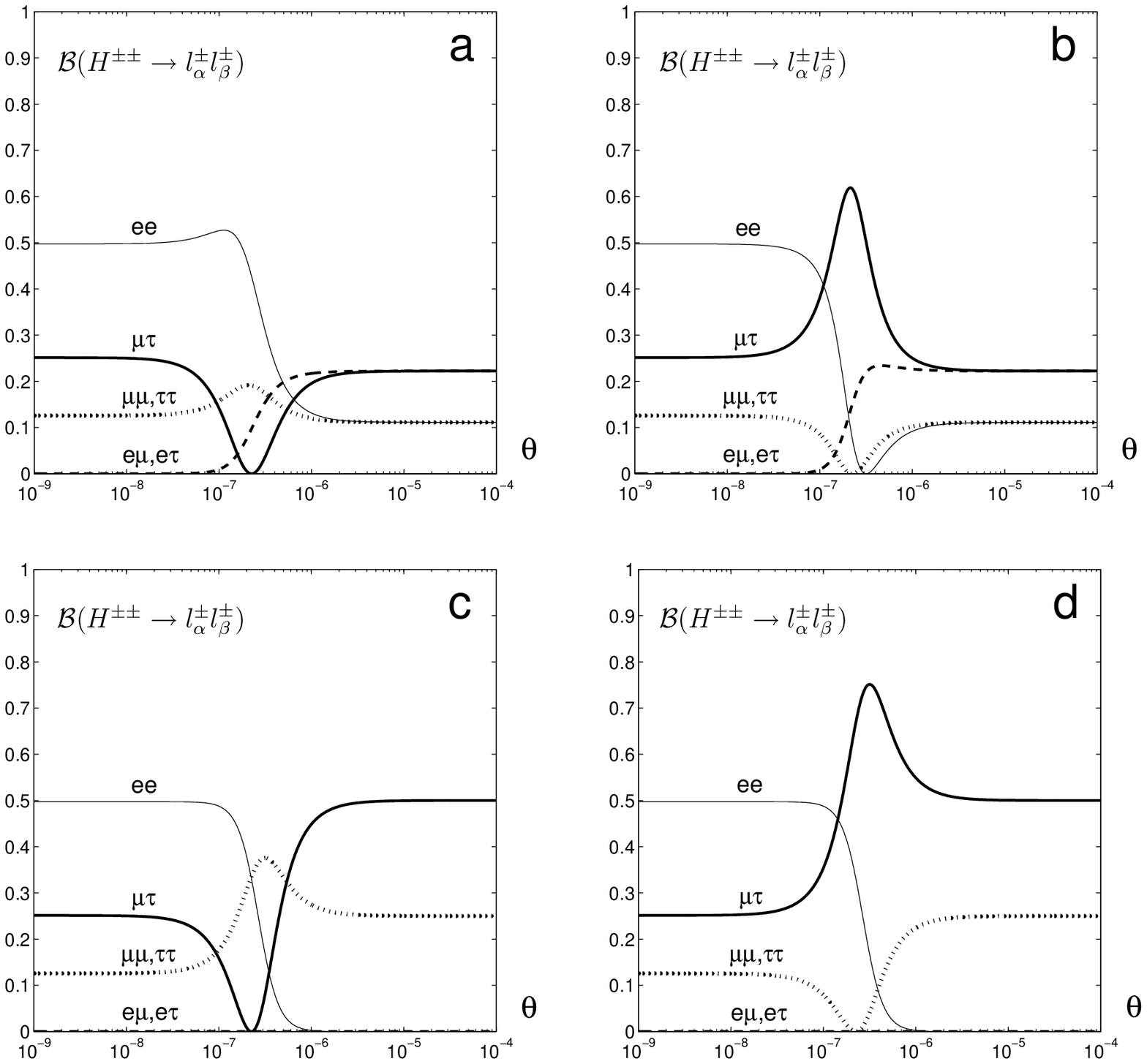, bbllx=2.5cm, bblly=19.6cm, bburx=6.6cm,
bbury=24.0cm, width=3.8cm, height=4cm, angle=0, clip=0}
\vspace{13.5cm} \caption{Branching ratios of $H^{\pm\pm} \to
l^\pm_\alpha l^\pm_\beta$ decays for the inverted hierarchy of
$m^{}_i$ with $m^{}_3 =0$:   (a) $\theta^{}_{14} = \theta^{}_{24} =
\theta^{}_{34} \equiv \theta$ and $\delta^{}_{14} = \delta^{}_{24} =
\delta^{}_{34} =0$;   (b) $\theta^{}_{14} = \theta^{}_{24} =
\theta^{}_{34} \equiv \theta$ and $\delta^{}_{14} = \delta^{}_{24} =
\delta^{}_{34} =\pi/2$;  (c) $\theta^{}_{14} = 0$, $\theta^{}_{24} =
\theta^{}_{34} \equiv \theta$ and $\delta^{}_{24} = \delta^{}_{34}
=0$; (d) $\theta^{}_{14} = 0$, $\theta^{}_{24} = \theta^{}_{34}
\equiv \theta$ and $\delta^{}_{24} = \delta^{}_{34} =\pi/2$. }
\end{figure}

%%%%%%%%%%%%%%%%%  figure 3      %%%%%%%%%%%%%%%%%%%%%%%%%%%%%%%%%%%%
\begin{figure}
\psfig{file=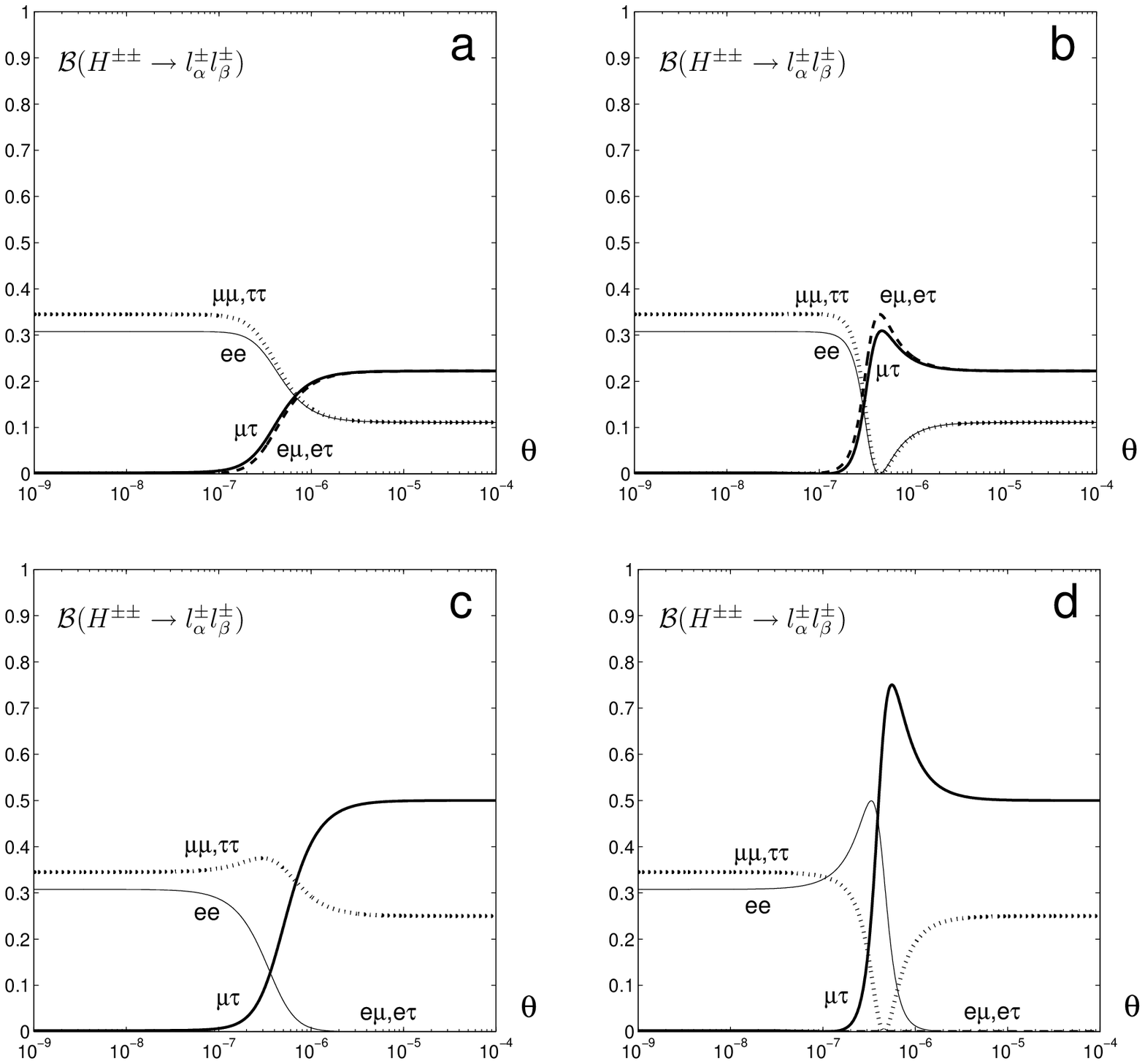, bbllx=2.5cm, bblly=19.6cm, bburx=6.6cm,
bbury=24.0cm, width=3.8cm, height=4cm, angle=0, clip=0}
\vspace{13.5cm} \caption{Branching ratios of $H^{\pm\pm} \to
l^\pm_\alpha l^\pm_\beta$ decays for the near degeneracy of $m^{}_i$
with $m^{}_3 > m^{}_2$:   (a) $\theta^{}_{14} = \theta^{}_{24} =
\theta^{}_{34} \equiv \theta$ and $\delta^{}_{14} = \delta^{}_{24} =
\delta^{}_{34} =0$; (b) $\theta^{}_{14} = \theta^{}_{24} =
\theta^{}_{34} \equiv \theta$ and $\delta^{}_{14} = \delta^{}_{24} =
\delta^{}_{34} =\pi/2$;  (c) $\theta^{}_{14} = 0$, $\theta^{}_{24} =
\theta^{}_{34} \equiv \theta$ and $\delta^{}_{24} = \delta^{}_{34}
=0$; (d) $\theta^{}_{14} = 0$, $\theta^{}_{24} = \theta^{}_{34}
\equiv \theta$ and $\delta^{}_{24} = \delta^{}_{34} =\pi/2$. }
\end{figure}


\begin{thebibliography}{99}

\bibitem{Venice} See, e.g., Z.Z. Xing, ``{\it Naturalness and Testability of
Seesaw Models at the LHC}", talk given at the IV International
Workshop on Neutrino Oscillations in Venice, April 15 - 18, 2008
(to appear in the proceedings).

\bibitem{PDG} Particle Data Group, W.M. Yao {\it et al.},
J. Phys. G {\bf 33}, 1 (2006).

\bibitem{Seesaw1} P. Minkowski, Phys. Lett. B {\bf 67}, 421 (1977);
T. Yanagida, in {\it Proceedings of the Workshop on Unified Theory
and the Baryon Number of the Universe}, edited by O. Sawada and A.
Sugamoto (KEK, Tsukuba, 1979); M. Gell-Mann, P. Ramond, and R.
Slansky, in {\it Supergravity}, edited by P. van Nieuwenhuizen and
D. Freedman (North Holland, Amsterdam, 1979); S.L. Glashow, in {\it
Quarks and Leptons}, edited by M. L$\acute{\rm e}$vy {\it et al.}
(Plenum, New York, 1980); R.N. Mohapatra and G. Senjanovic, Phys.
Rev. Lett. {\bf 44}, 912 (1980).

\bibitem{Seesaw2} M. Magg and C. Wetterich, Phys. Lett. B {\bf 94}, 61 (1980);
J. Schechter and J.W.F. Valle, Phys. Rev. D {\bf 22}, 2227 (1980);
T.P. Cheng and L.F. Li, Phys. Rev. D {\bf 22}, 2860 (1980); G.
Lazarides, Q. Shafi, and C. Wetterich, Nucl. Phys. B {\bf 181},
287 (1981); R.N. Mohapatra and G. Senjanovic, Phys. Rev. D {\bf
23}, 165 (1981).

\bibitem{Gu} P.H. Gu, H. Zhang, and S. Zhou, Phys. Rev. D {\bf
74}, 076002 (2006); A.H. Chan, H. Fritzsch, S. Luo, and Z.Z. Xing,
Phys. Rev. D {\bf 76}, 073009 (2007).

\bibitem{Zhou} W. Chao, Z. Si, Z.Z. Xing, and S. Zhou,
arXiv:0804.1265 [hep-ph].

\bibitem{Triplet} See, e.g., K. Huitu, J. Maalampi, A. Pietila, and M. Raidal,
Nucl. Phys. B {\bf 487}, 27 (1997); B. Dion {\it et al.}, Phys. Rev.
D {\bf 59}, 075006 (1999); E.J. Chun, K.Y. Lee, and S.C. Park, Phys.
Lett. B {\bf 566}, 142 (2003); A.G. Akeroyd and M. Aoki, Phys. Rev.
D {\bf 72}, 035011 (2005); A. Hektor {\it et al.}, Nucl. Phys. B
{\bf 787}, 198 (2007); T. Han, B. Mukhopadhyaya, Z. Si, and K. Wang,
Phys. Rev. D {\bf 76}, 075013 (2007); C.S. Chen, C.Q. Geng, and D.V.
Zhuridov, arXiv:0801.2011.

\bibitem{KS} W.Y. Keung and G. Senjanovic, Phys. Rev. Lett. {\bf
50}, 1427 (1983).

\bibitem{LL} See, e.g.,
J. Garayoa and T. Schwetz, arXiv:0712.1453; M. Kadastik, M.
Raidal, and L. Rebane, arXiv:0712.3912; A.G. Akeroyd, M. Aoki, and
H. Sugiyama, arXiv:0712.4019; P. Fileviez P\'{e}rez, T. Han, G.Y.
Huang, T. Li, and K. Wang, arXiv:0803.3450; Z.Z. Xing,
arXiv:0805.0416.

\bibitem{Han2} For a systematic study of how to test the triplet
seesaw model at the LHC, see: P. Fileviez P\'{e}rez, T. Han, G.Y.
Huang, T. Li, and K. Wang, arXiv:0805.3536.

\bibitem{Xing08} Z.Z. Xing, Phys. Lett. B {\bf 660}, 515 (2008).

\bibitem{FX01} H. Fritzsch and Z.Z. Xing, Phys. Lett. B {\bf 517},
363 (2001); Z.Z. Xing, Int. J. Mod. Phys. A {\bf 19}, 1 (2004).

\bibitem{Vissani} A. Strumia and F. Vissani, hep-ph/0606054; G.L.
Fogli {\it et al.}, arXiv:0805.2517.

\bibitem{Antusch} S. Antusch, C. Biggio, E. Fernandez-Martinez,
M.B. Gavela, and J. Lopez-Pavon, JHEP {\bf 0610}, 084 (2006); E.
Fernandez-Martinez, M. B. Gavela, J. Lopez-Pavon, and O. Yasuda,
Phys. Lett. B {\bf 649}, 427 (2007); W. Chao, S. Luo, Z.Z. Xing, and
S. Zhou, Phys. Rev. D {\bf 77}, 016001 (2008); S. Luo,
arXiv:0804.4897.

\bibitem{Han} T. Han and B. Zhang, Phys. Rev. Lett. {\bf 97},
171804 (2006).

\bibitem{TB} P.F. Harrison, D.H. Perkins, and W.G. Scott, Phys.
Lett. B {\bf 530}, 167 (2002); Z.Z. Xing, Phys. Lett. B {\bf 533},
85 (2002); P.F. Harrison and W.G. Scott, Phys. Lett. B {\bf 535},
163 (2002); X.G. He and A. Zee, Phys. Lett. B {\bf 560}, 87 (2003).

\bibitem{Chao} W. Chao, Z.Z. Xing, and S. Zhou, in preparation.

\end{thebibliography}
\end{document}